\documentclass[useAMS,usenatbib,usegraphicx]{mn2e}
\usepackage{graphicx}
\usepackage{color}

\newcommand{\mincir}{\raise
-3.truept\hbox{\rlap{\hbox{$\sim$}}\raise4.truept\hbox{$<$}\ }}
\newcommand{\magcir}{\raise
-3.truept\hbox{\rlap{\hbox{$\sim$}}\raise4.truept\hbox{$>$}\ }}
\usepackage{morefloats/morefloats}
\usepackage{times}
\usepackage{rotate} 
\usepackage{amssymb} 
\usepackage{rotating}
\usepackage{aas_macros}

\title[X-ray AGN Clustering]{Clustering, Bias and the Accretion Mode of X-ray selected AGN}
\author[Koutoulidis et al.]  {L. Koutoulidis$^{1,2}$,
  M. Plionis$^{1,3}$, I.Georgantopoulos$^{1,4}$, N. Fanidakis$^{5}$\\
  $^1$ National Observatory of Athens,  V.  Paulou \&  I.  Metaxa,
  11532, Greece\\
  $^2$ Phyiscs Dept., University of Patras, Greece \\
  $^3$ Instituto Nacional de Astrof\'isica, \'Optica y Electr\'onica,
  L.Enrique Erro 1, Tonantzintla, Puebla, M\'exico C.P. 72840 \\
  $^4$ INAF-Osservatorio Astronomico di Bologna, Via Ranzani 1, 40127, Italy\\ 
$^5$ Max Planck Institute for Astronomy, K\"onigstuhl 17 D-69117 Heidelberg, Germany}

\begin{document}
\maketitle

\begin{abstract}
We present the spatial clustering properties of 1466 X-ray selected
AGN compiled from the {\em Chandra} CDF-N, CDF-S, eCDF-S, COSMOS and
AEGIS fields in the $0.5-8$ keV band. The X-ray sources span the redshift
interval $0<z<3$ and have a median value of $\bar{z}=0.976$.
 We employ the projected two-point correlation function to infer the
 spatial clustering and find a clustering length of $r_0= 7.2\pm
 0.6 h^{-1}$ Mpc and a slope of $\gamma=1.48\pm 0.12$, which
 corresponds to a bias of $b(\bar{z})=2.26 \pm 0.16$. 
Using two different halo bias models, we consistently estimate
an average dark-matter host halo mass of $M_{\rm h}\simeq 1.3 (\pm 0.3)
\times 10^{13} h^{-1} M_\odot$. The X-ray AGN bias and the corresponding 
dark-matter host halo mass, are significantly higher than the corresponding
values of optically selected AGN (at the same redshifts). %indicating different populations of AGN.
The redshift evolution of the X-ray selected AGN bias indicates, in
agreement with other recent studies, that a unique dark-matter halo
mass does not fit well the bias at all the different redshifts probed.
 Furthermore, we investigate if there is a dependence of the clustering
strength on X-ray luminosity. To this end we consider only 650 sources
around $z\sim 1$ and we apply a procedure to disentangle the 
dependence of clustering on redshift. 
We find indications for a positive dependence
of the clustering length on X-ray luminosity, in the sense that the more
luminous sources have a larger clustering length and hence a higher 
dark-matter halo mass. In detail we find for an average luminosity
difference of $\delta\log_{10} L_x\simeq 1$ a halo mass difference of
a factor of $\sim$3.

These findings appear to be consistent with a galaxy-formation model
where the gas accreted onto the supermassive black hole in
intermediate luminosity AGN comes mostly from the hot-halo atmosphere
around the host galaxy.
\end{abstract}

\section{Introduction}\label{sec_intro}
One of the most remarkable astonomical findings of the 
last decade is the discovery of  the scaling relations between the mass of supermassive 
black holes (BHs) and the properties of the large-scale environment 
of their host galaxies. In particular, observations suggest that there 
is a tight correlation between the BH mass and the bulge velocity 
dispersion, stellar mass and luminosity of the spheroidal component 
of the host galaxy \cite[e.g.,][]{{Ferrarese, Magorrian}}, \cite[][]{Haering, 
Gultekin}. 
These observational relations indicate that the cosmic growth of BH mass 
is strongly coupled to the evolution of the host spheroid. However, the 
physical mechanism that shapes these relations is still unknown.

In most semi-analytic models of galaxy formation 
\citep[e.g.,][]{Croton06, Bower06, Lagos08}, \cite{Somerville08}
it is assumed that the active galactic nucleus (AGN) associated with
 accreting black hole regulates the formation of new stars.
The mechanisms adopted by semi-analytical models for triggering
AGN activity include major galaxy mergers for the most luminous
AGN \citep{DiMatteo05, Hopkins06, Marulli09}, while it is
possible that in the lowest luminosity AGN regime secular disk
instabilities or minor interactions play the key role 
\citep{HopkinsHernquist06, Bournaud11}.
These different AGN fueling modes  make diverse predictions for the environment of
the galaxies that host AGN \citep{Shankar09}.  
For example,  in the major-merger
scenario, only weak luminosity dependence on clustering is
expected \citep{Hopkins05, Lidz06, Bonoli09}. Hence, 
observational studies of AGN clustering and its dependence on luminosity
can place valuable constraints on the AGN fueling modes and consequently 
on the AGN--galaxy co-evolution models.

The clustering of AGN has been studied with excellent 
statistics mainly in the optical and particularly in large area surveys,
such as the 2QZ \citep[2dF QSO Redshift Survey][]{Croom05,Porciani06}
and the SDSS \citep[Sloan Digital Sky
Survey,][]{Li06,Shen09,Ross09}. These surveys found no evidence
for a strong dependence of clustering on luminosity
 \citep[e.g.,][]{Croom05, Myers06, DaAngela08}
 although \citet{Shen09} detect an excess of clustering for their
 10\% brightest quasars.   
However, optical QSO may represent only the tip of the iceberg
of the AGN population. Very deep X-ray surveys find a surface density of about
10,000 deg$^{-2}$ \citep{Xue11},  which is about two orders of
magnitude higher than that found in optical QSO surveys \citep{Wolf03}. 
Several studies have explored the angular clustering of AGN  in X-ray
wavelengths
\citep[][]{Vikl98,Akylas,Yang03}, \citep{Basilakos04, Basilakos05}, 
\cite{Gandhi}, \cite{Pucc}, \cite{Carrera},
\cite{Miyaji07}, \cite{Plionis08}, \cite{Ebrero09}, \cite{Elyiv12}. 
These studies measure the projected
angular clustering and then via Limber's equation \citep{Peebles1980} 
derive the corresponding spatial clustering length. However,
in this method an a priori knowledge of the redshift distribution
$dN/dz$ is needed and thus, the uncertainties  may be
appreciable. \citet{Plionis08} first reported strong indications
for a luminosity dependent clustering of X-ray AGN in the CDF fields
for both soft and hard bands in the sense that X-ray luminous AGN  lie
in more massive dark-matter (DM) halos compared to the less luminous ones. 

 Recently, several studies have attempted to measure the spatial
 correlation function of X-ray selected AGN 
by employing spectroscopic redshifts
 to estimate their correct distances
\citep{Mulli04, Gilli05, Yang06}, \citep{Gilli09, Hickox09, Coil09},
\cite{Krumpe10}, \citep{Cappelluti10, Miyaji11}, 
\citep{Starikova11, Allevato11}.
These studies show that X-ray AGN are typically hosted in DM
 halos with mass of the order of $12.5 <\log_{10} (M_{\rm
  h}/[h^{-1}M_\odot])< 13.5$, at a redshift $z\sim1$.
On the issue of luminosity dependent clustering the results are contentious. 
\citet{Yang06} did not detect strong correlation between X-ray
luminosity and the clustering amplitude for {\it
Chandra} sources. \citet{Gilli09} performed a similar investigation 
in the XMM-COSMOS field dividing the sample below
and above $L_x (0.5-10 {\rm keV}) = 10^{44}~\rm{erg~s}^{-1}$
and did not find any luminosity dependence. Similarly,
\citet{Starikova11}, using 1282 Chandra/Bootes sources, did not find
a significant dependence of clustering on luminosity.

On the other hand, \citet{Coil09} using AGN in the redshift range $z=0.7-1.4$
from the AEGIS field found a weak evidence for a luminosity
dependent clustering, but not at a statistically significant level. In addition, 
\citet{Krumpe10} using low redshift ($z\sim0.25$) AGN, selected from
the {\it ROSAT} all-sky survey and cross matched with the SDSS, found
that $L_x > 10^{44}~\rm{erg~s}^{-1}$ sources are clustered more strongly
than lower luminosity sources.
In another recent study, \citet{Cappelluti10} using a sample
of 199 Swift-BAT sources in the $15-55~\rm{keV}$ band, found a
marginally significant luminosity dependent clustering.

In this paper we use a sample of $1466$ sources
selected in the $0.5-8~\rm{keV}$ band from a variety of deep {\it Chandra} 
X-ray surveys, to estimate the spatial correlation function 
and typical DM host halo mass of X-ray selected AGN. Our aim is to investigate
whether there is a luminosity dependence on clustering, but also to obtain insight 
into the black-hole fueling mechanisms, using the largest X-ray selected AGN 
sample used so far for this scope. The paper is organized as follows. 
In Section~2, we briefly discuss
the X-ray surveys we consider in this paper and in Section~3, we describe the 
methodology of our clustering analysis. In Section 4, we present our findings for the 
AGN correlation function as estimated by the joint and individual X-ray samples.
In Section~5, we calculate the bias of AGN and its redshift evolution,
and estimate the average mass of the DM halos hosting X-ray
AGN. In Section~6, we investigate 
the dependence of the halo mass on X-ray luminosity and compare our findings
with theoretical galaxy formation models for the evolution of accreting BHs.
Finally, we present in Section~7 our conclusions.
For comparison reasons with previous works we adopt, unless otherwise stated, 
a flat cosmology with a present-day matter density parameter $\Omega_{m} = 0.3$,
a cosmological constant $\Omega_{\rm{\Lambda}}=0.7$ and baryon density 
$\Omega_{b}=0.04$. 
%\textbf{(WMAP7 cosmology? what is the $\sigma_8$)}. 
The Hubble constant is expressed in units of $h$ as $H_{0} = 100~h$ km
s$^{-1}$ Mpc $^{-1}$.
% (distances and luminosities are parametrized by $h^{-1}$ Mpc).

\section{AGN CATALOGS}
We make use of X-ray data coming from the five deepest X-ray surveys,
namely the Chandra Deep Field South and North (CDF-S and CDF-N), the
AEGIS, the extended Chandra Deep Field South (ECDF-S) and the COSMOS survey.
These cover a variety of exposures (from $4~\rm{Ms}$ down to $40~\rm{ksec}$) 
and surveyed area and thus, cover extensively the luminosity-redshift
space. Moreover, these fields contain excellent quality spectroscopic
observations and therefore good quality redshifts, which are essential in
this project. Below we present briefly the main characteristics of the
X-ray surveys used in this work.

\subsection{CHANDRA DEEP FIELD NORTH}
The deep pencil beam CDF-N survey covers an area of $448~\rm{arcmin}^2$,
is centered at {\rm$a=12^h 36^m 49^s, \delta=+62^\circ 12^, 58^{,,}$}
and consists of 20 individual ACIS-I (Advanced CCD Imaging
Spectrometer) pointings. The combined observations provide the deepest
X-ray sample currently available together with the CDF-S. 
Here, we use the X-ray $2~\rm{Ms}$ source catalogue of
\citet{Alexander2003}, 
with a sensitivity of $\sim 10^{-17}$ erg cm$^{-2}$ s$^{-1}$, which
consists of 503 sources in the $0.5-8$ keV band. Spectroscopic
redshifts were used for 243 X-ray sources in the redshift interval
$z=0-3$ from \citet[][and references therein]{Trouille08}.

\subsection{CHANDRA DEEP FIELD SOUTH}
The deep pencil beam CDF-S survey covers an area of 436 arcmin$^2$
and the average aim point is $\rm{a=03^h32^m28^s8,
\delta=-27^\circ48^,23^{,,} (J2000)}$. The analysis of all 23
observations is presented in \citep{Luo08}. We use the 2Ms X-ray
source catalogue of \citet{Luo08}, which consists of 462 X-ray
sources. Spectroscopic redshifts were used for 219 X-ray sources in
the redshift interval {\rm$z=0-3$} from \citet{Luo10}.

\subsection{AEGIS}
The ultra deep field survey comprises of pointings at 8 separate
positions, each with nominal exposure of $200$~ks, covering a total area of
approximately $0.67~\rm{deg}^2$ centered at {\rm$a=14^h17^m,
  \delta=+52^\circ30^,$} in a strip of length $2~\rm{deg}$. The flux
limit of the survey is $S\sim 10^{-16}$ erg s$^{-1}$cm$^{-2}$ in the full band. We use
the X-ray source catalogue of \citet{Laird09}, which consists of 1325
sources. Spectroscopic redshifts were used for 392 X-ray sources in
the redshift interval {\rm$z=0-3$} from DEEP2 \citep{Davis01,Davis03,Coil09}.

\subsection{COSMOS}
The Chandra COSMOS Survey covers the central $0.5~\rm{deg}^2$ 
area of the COSMOS field with an effective exposure of
$\thicksim160\rm{ks}$ and the rest of the field with an effective exposure of
$\thicksim80\rm{ks}$. The limiting source detection depths are 
 $5.7 \times 10^{-16}$ erg s$^{-1}$ cm$^{-2}$ in the full band. We use the source
catalog of \citet{Elvis09}, which consists of 1761
sources. Spectroscopic redshifts were used for 417 X-ray sources from
\citet{Brusa10}. %We must notice that we checked which angular
%separation is secure in order to take care that one source of Chandra
%corresponds to one source of XMM, due to different resolution.

\subsection{EXTENDED CHANDRA DEEP FIELD SOUTH}
The ECDF-S survey consists of 4 Chandra $250~\rm{ks}$ ACIS-I pointings
covering $\sim 0.3$ deg$^2$ and surrounding the original
CDF-S. Source detection has been performed by \citet{Lehmer05} and
\citet{Virani06}. Here we use the source detection of \citet{Lehmer05} 
in which $762$ sources have been detected. The flux limit of
ECDF-S is $S\sim 10^{-16}$ erg s$^{-1}$cm$^{-2}$. Spectroscopic
redshifts were used for $288$ sources in the redshift interval $z=0-3$ from
\citet{Silverman10}. When combining all fields, we exclude
sources that are detected both in ECDF-S and CDF-S and keep only
those detected in the ECDF-S.

 \section{METHODOLOGY}

\subsection{Theoretical Considerations}
The main statistic used to measure the clustering of extragalactic sources is
the two-point correlation function {\rm$\xi(r)$}. {\rm$\xi(r)$} describes the
excess probability over random of finding a pair with one object in
an elemental volume {\rm$dV_1$} and the second in the elemental volume {\rm$dV_2$},
separated by a distance $r$ \citep[e.g.,][]{Peebles1980}. Its mathematical
description is given by $dP = \langle n\rangle^2 [1+ \xi(s)] dV_1dV_2$, where $\langle
n\rangle$ is the mean space density of the sources under study and $s=
c z$.
When measuring $\xi$ directly from redshift catalogues of sources, we 
include the distorting effect of peculiar velocities, since the true
distance of a source is $r=(s-{\bf v}_{p}\cdot{\bf r})/H_0$, where ${\bf
 v}_p\cdot {\bf r}$ is the component of the peculiar velocity of the 
 source along the line of sight. 
In order to avoid such effects one can either measure the angular
clustering, which is not hampered by the effects of $z$-distortions, and
then, under some assumptions, infer the spatial correlation
function through the Limber's integral equation \citep{Limber1953}. 
Alternatively one can use the redshift information to
measure the so-called projected correlation
function, $w_p(r_p)$ \citep[e.g.,][]{DavisPeeb1983} and then infer the
spatial clustering. 

To this end,
one deconvolves the redshift-based distance of a source, $s$, in two
components, one parallel ($\pi$) and one perpendicular ($r_p$) to
the line of sight, i.e., $s=(r_p^2+\pi^2)^{1/2}$, and thus the redshift-space 
correlation function can be written as $\xi(s)=\xi(r_p, \pi)$.
Since redshift space distortions affect only the $\pi$
component, one can estimate the free of $z$-space distortions
projected correlation function, $w_p(r_p)$, 
by integrating $\xi(r_p,\pi)$ along $\pi$:
\begin{equation}\label{eq:wp}
w_p(r_p)=2\int_{0}^{\infty}\xi(r_p,\pi)\mathrm{d}\pi.
\end{equation}

Once we estimate the projected correlation function, $w_p(r_p)$, we can
recover the real space correlation function, since the two are related
according to \citep{DavisPeeb1983}
\begin{equation}\label{eq:wp}
 w_p(r_p)=2\int_{0}^{\infty}\xi(\sqrt{r_p^2+\pi^2}) {\rm d}\pi =2\int_{r_p}^{\infty}
 \frac{r\xi(r)\mathrm{d}r}{\sqrt{r^2-r_p^2}}\;.
 \end{equation}
Modelling $\xi(r)$ as a power law, $\xi(r)=\left(r/r_0\right)^{-\gamma}$
%following Davis\&Peebles (1983) calculated this integral numerically
one obtains,
\begin{equation}\label{eq:wp_model}
w_p(r_p)=A_\gamma r_p \left(\frac{r_0}{r_p}\right)^{\gamma},
\end{equation}
with
\begin{equation}
A\gamma=\Gamma\left(\frac{1}{2}\right)
\Gamma\left(\frac{\gamma-1}{2}\right)/\Gamma\left(\frac{\gamma}{2}\right),
\end{equation} 
where $\Gamma$ is the usual gamma function.

However, it should be noted that although Eq.~(\ref{eq:wp_model}) strictly holds for
$\pi_{\rm max}=\infty$, practically we always impose a cutoff
$\pi_{\rm max}$ (for reasons discussed in the next subsection). This
introduces an underestimation of the underlying correlation
function, which is an increasing function of separation $r_p$. 
For a power law correlation function this underestimation is easily
inferred from Eq.~(\ref{eq:wp}) and is given by \citep[e.g.,][]{Starikova11}
\begin{equation}
C_{\gamma}(r_p)=\frac{\int_0^{\pi_{\rm max}} (r^2-\pi^2)^{-\gamma/2} d\pi}
{\int_0^\infty (r^2-\pi^2)^{-\gamma/2} d\pi}\;.
\end{equation}
Thus, by taking into account the above statistical correction, and under the assumption
of the power-law correlation function,
one can recover the corrected spatial correlation function, $\xi(r_p)$, from
the fit to the measured $w_p(r_p)$ according to  (which provides also
the value of $\gamma$):
\begin{equation}\label{eq:corr}
\xi(r_p)=\frac{1}{A_\gamma C_{\gamma}(r_p)} \frac{w_p(r_p)}{r_p} \;.
\end{equation}
However, at large separations the correction factor increasingly
dominates over the signal and thus it constitutes the correction
procedure unreliable. 
Alternatively, as can be easily shown using Eqs. (\ref{eq:wp_model})
and (\ref{eq:corr}), a crude estimate of the corrected spatial
correlation length can be provided by:
\begin{equation}\label{eq:crude}
r_{o,c}=r_0 C_\gamma(r_0)^{-1/\gamma}\;,
\end{equation}
where $r_0$ and $\gamma$ are derived from fitting the data to
Eq.~(\ref{eq:wp_model}). 

\subsection{Correlation Function Estimator}
As a first step we calculate $\xi(r_p,\pi)$ using the estimator \citep[cf.,][]{Kerscher}
\begin{equation}\label{eq:xi}
\xi(r_p,\pi)=\frac{N_R}{N_D}\frac{DD(r_p,\pi)}{DR(r_p,\pi)}\;,
\end{equation} 
where $N_D$ and $N_R$ are the number of data and random sources,
respectively, while $DD(r_p,\pi)$ and
$DR(r_p,\pi)$ are the number of data-data and data-random pairs,
respectively. We then estimate the redshift-space correlation function, $\xi(s)$, in the
range $s=0.16-40~h^{-1}$ Mpc and the projected correlation function,
$w_p(r_p)$, in the separation range $r_p=0.16-20~h^{-1}$ Mpc.
Note that large separations in the $\pi$ direction add mostly noise to
the above estimator and therefore the integration is truncated for
separations larger than $\pi_{\rm max}$. The choice of  $\pi_{\rm max}$ is a compromise
between having an optimal signal to noise ratio for $\xi$ and reducing the
excess noise from high $\pi$ separations. The majority of studies in the
literature usually assume $\pi_{\rm max} \in [5, 30]~h^{-1}$ Mpc.

The correlation function uncertainty is estimated according to
\begin{equation}
\sigma_{w_p}=\sqrt{3} (1+w_p)/\sqrt{DD}\;,
\end{equation}
which corresponds to that expected by the bootstrap technique
\citep{Mo92}.
In this work, we bin the source pairs in logarithmic intervals of
 $\delta\log_{10}(r_p,\pi)=0.15$ and $\delta\log_{10}(s)=0.17$ for the $w_p(r_p)$
and $\xi(s)$ correlation functions respectively.
Finally, we use a $\chi^2$ minimization 
procedure between data and the power-law
model for either type of the correlation function to derive the best fit $r_0$
and {\rm$\gamma$} parameters. We carefully choose the range of separations  
in order to obtain the best  power-law fit to the data and we impose a lower
separation limit of $r_p\sim 1.5~h^{-1}$ Mpc to minimize non-linear effects.

\begin{figure}
\begin{center}
\includegraphics[width=4.1cm]{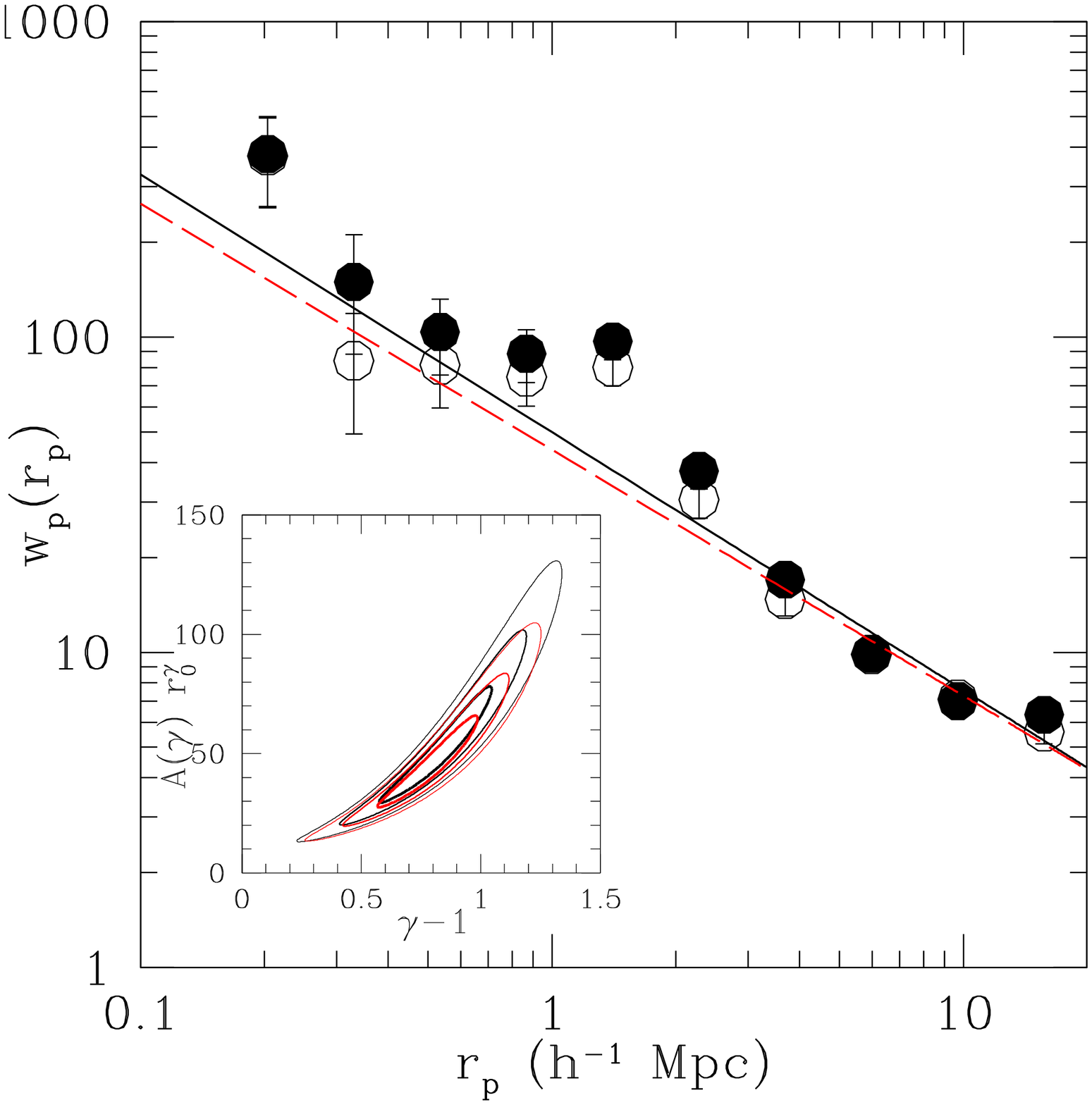} \hfill
\includegraphics[width=4.1cm]{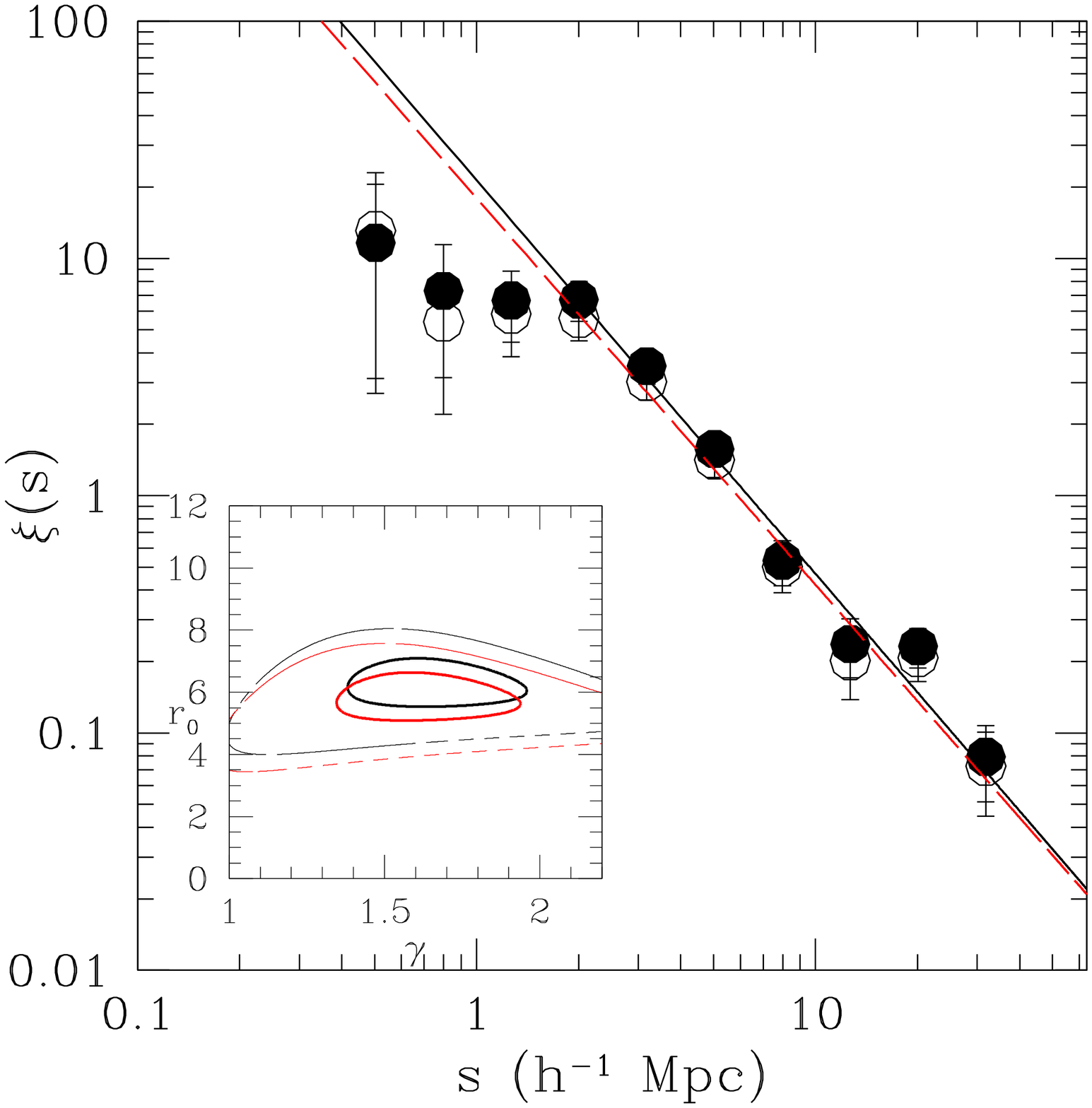} 
\end{center}
\caption{Comparison of the correlation function results based on the
  G05 random construction method (filled symbols) with that of the standard
  `sensitivity map' method (empty symbols), applied on the AEGIS
  full band (0.5-8 keV) catalog. 
  Uncertainties correspond to {\rm $1\sigma$} Poisson errors. 
  The best fit power
  laws are shown as black lines for the G05 method and as dashed red lines
  for the sensitivity map method.
  In the inset panels we show
  the likelihood {\rm$1\sigma, 2\sigma, 3\sigma$} contours in the
  2-parameter solution space. 
{\em Right Panel}: Redshift space correlation function, $\xi(s)$.
{\em Left Panel}: Projected correlation function, $w_p(r_p)$.}
\label{agil}
\end{figure}

\subsection{Random Catalogue Construction}
To estimate the spatial correlation function of a sample of sources
one needs to construct a large mock comparison sample with a random spatial
distribution within the survey area, which also reproduces all the
systematic biases that are present in the source sample (i.e.,
instrumental biases due to the Point Spread Function variation,
vignetting, etc). Also special care has to be taken to reproduce any
biases that enter through the optical counterpart spectroscopic
observations strategy (cf., due to the positioning of the masks within
the field of view and of the slits within the masks, etc).

To this end we will follow the random catalogue construction procedure
of \citet[][hereafter G05]{Gilli05}, which is based on reshuffling
only the source redshifts, smoothing the corresponding redshift distribution, while
keeping the angular coordinates unchanged. We will test the efficiency
of this method by comparing the outcome correlation function results
with those of the standard method that takes into account the details
of the sensitivity maps of each XMM pointing. We further 
test the two methods using N-body simulations for the underlying DM density field.

\subsubsection{Random redshift reshuffling method} 
In order to assign random redshifts to the mock sample, the 
source redshift distribution is smoothed  using a Gaussian
kernel with a smoothing length of $\sigma_z =0.3$.
This offers a compromise between scales
that are either too small, and thus may reproduce the $z$-space
clustering, or too large and thus over-smooth the observed
redshift distribution.
We verified that our results do not change significantly when using a
smoothing length in the range $\sigma_z =0.2-0.4$.

\subsubsection{Standard `sensitivity  map' method}
According to the standard method of producing random catalogs, each
simulated source is placed at a random position on the sky,
with a flux randomly extracted from the observed source $\log N-\log
S$ (source number-flux distribution).
If the flux is above the value allowed by the sensitivity map at that
position, the simulated source is kept in the random sample and a
random redshift is also assigned to it from the observed source redshift
distribution $N(z)$ (optimally taking into account the variation of 
$N(z)$ as a function of flux). 
The disadvantage of this method is that it does not take
into account any unknown inhomogeneities and systematics of the follow-up
spectroscopic observations.

\subsubsection{Testing the efficiency of the two methods}
In order to test the efficiency of the two random catalogue construction
methods we use the AEGIS survey, consisting of 8 fields. To construct
the standard method random catalogue we use the sensitivity maps of
\citep{Laird09} and we compute the projected and redshift-space
2-point correlation functions.
The best fit correlation length found is
$r_0= 4.28 \pm 0.14~h^{-1}$ Mpc for $\gamma=1.8$ which is in
excellent agreement with the results based on the G05 method:
$r_0= 4.13 \pm 0.14~h^{-1}$ Mpc. In Fig.~\ref{agil} we compare the
correlation functions for the two random construction methods.

\subsubsection{A further test using simulations}
We use a 500 $h^{-1}$ Mpc cube simulation 
of a flat $\Omega_m=0.3$ and $\sigma_8=0.9$ cosmology with $512^3$
particles \citep{Ragone07}, 
to further test the robustness of the G05 random sample construction method.
In order to have a relatively wide redshift range we replicate randomly the box to get
 an effective volume of $1500^{3}~h^{-3}$ Mpc$^{3}$. To obtain a
 relatively large but manageable source density, we selected DM
  halos with $M_{\rm h}> 1.5 \times 10^{14}~h^{-1} M_{\odot}$, which resulted in
 a total number of haloes within our large simulation volume of $N= 9\times 2312$.

We estimate the actual halo correlation function according to the
{\em direct} method, which entails counting the number of DM halo
pairs within spherical shells around a given halo. The
corresponding number of random pairs is
estimated by $\langle n \delta\rangle V$, with
$\langle n \rangle$ the mean number density of halos in the whole
simulation volume and $\delta V$
 the volume of the spherical shell i.e. $4/3\pi \delta r^3$, with
 $\delta r$ being the width of the spherical shell. 

We also estimate the halo correlation function using the G05 method,
for which we transform the halo Cartesian coordinates into
spherical coordinates, determining also their distance distribution.
The clustering results of the two methods are in excellent agreement
%are shown in Table Fig.~\ref{red_meth}, 
%from which it becomes evident that they are in very good agreement, providing
providing
 $\gamma=1.72 (\pm0.05)$ and $1.68 (\pm0.12)$ as well as 
$r_0=13.9(\pm 0.3)$ and $13.2(\pm 0.5)~h^{-1}$ Mpc, for the {\em direct} and
G05 correlation function estimations, respectively. 

\begin{figure}
\includegraphics[width=8.1cm]{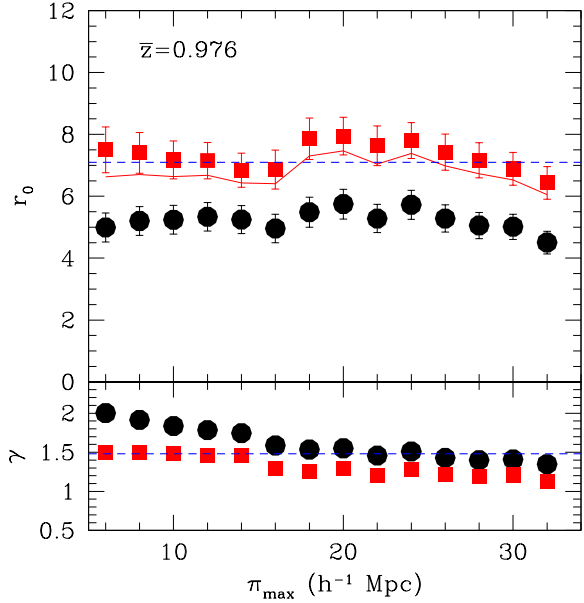}
\caption{The resulting amplitude $r_0$, and slope $\gamma$, of the
  correlation function analysis as a function of the value of the
  $\pi_{\rm max}$
  cutoff. The filled circles correspond to the results based on $w_p(r_p)$, while
  the red squares on the results of $\xi(r_p)$. The continuous red line
  corresponds to the crude estimation of the corrected  spatial  amplitude
  based on Eq.~(\ref{eq:crude}), while the dashed blue line correspond
  to the finally adopted
  results (which corresponds to the $\pi_{\rm max}=10~h^{-1}$ Mpc case).  }
\label{fig:pmax}
%The lower panel shows the reduced $\chi^2$ value of the power-law fit
%to the observed correlation function.}
\end{figure}

\begin{figure}
\includegraphics[width=8.1cm]{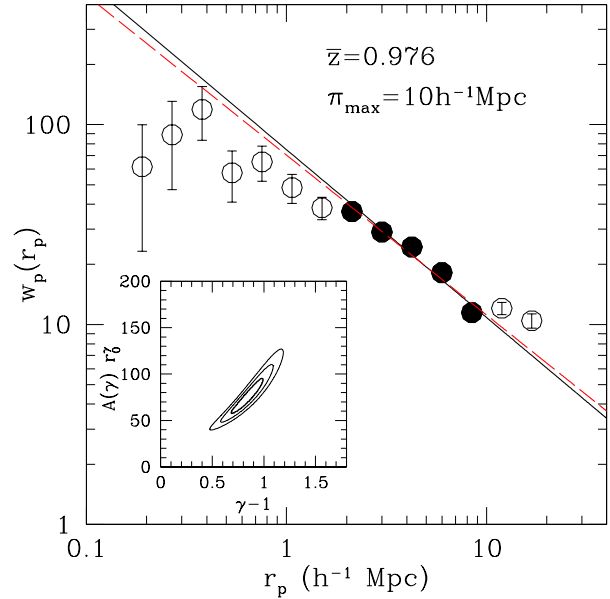} \vfill
\includegraphics[width=8.1cm]{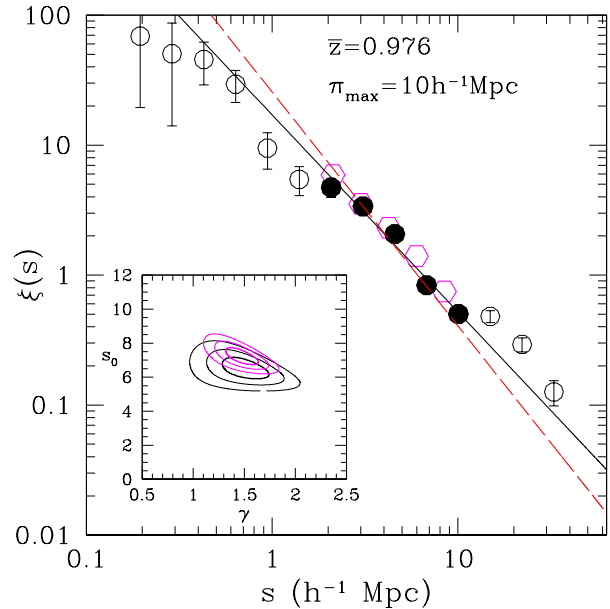} 
\caption{Correlation functions of the joint X-ray point-source sample: 
The projected correlation function, $w_p(r_p)$ is shown in the upper panel, and
the $z$-space correlation function, $\xi(s)$, is shown in the lower panel.
The filled points indicate the range over which a power-law fit was applied 
(black line corresponds to a fit with free $\gamma$, while the red line
to that for $\gamma=1.8$). 
The real-space correlation function, $\xi(r_p)$ (via Eq.~\ref{eq:corr}) is
indicated in the lower panel by magenta pentagons. The inset panels show
the 1, 2 and 3$\sigma$ likelihood contours in the 2-parameter plane of
the power-law fit solutions.}
\label{fig:3}
\end{figure}

\section{RESULTS}

 \subsection{The Joint X-ray Sample}
Combining all five fields, we obtain in total 1466 X-ray sources with
spectroscopic redshifts. The median redshift of the source sample is 
$\bar{z}=0.976$. In order to estimate the joint sample
correlation function we generate random catalogs for
each field separately and then combine them together.
Note that we take care to count only once the sources that are common
in both the CDF-S and eCDF-S fields.  

We remind the reader that we fit a power-law to the measured
correlation function over a range of separations for which
such a model represents well the data. In order to avoid non-linear
effects we use a lower separation limit, $r_{p} \simeq
1.5~h^{-1}$ Mpc, for the fit of $s$. We also impose an upper separation limit 
($\simeq 10~h^{-1}$ Mpc) since we find a change above this scale in the slope of both
projected and $z$-space correlation functions.
% and therefore we use as an upper separation limit for our
%power-law fits the scale $\simeq 10~h^{-1}$ Mpc.

We now investigate the sensitivity of the correlation function on
the cutoff separation along the line of sight, $\pi_{\rm max}$, which we vary in the
range $[5, 30]~h^{-1}$ Mpc. For $\pi_{\rm max}> 30~h^{-1}~\rm{Mpc}$ the noise introduced by
uncorrelated pairs reduces significantly the correlation function amplitude and slope. In
Fig.~(\ref{fig:pmax}) we show how the amplitude $r_0$ 
and the slope $\gamma$ of the projected correlation
function, $w_p(r_p)$ (open circles), and the corresponding real-space correlation
function, $\xi(r_p)$ (red squares), depend on $\pi_{\rm max}$. 
It is clear that both $r_0$ and $\gamma$ are
relatively constant in the investigated $\pi_{\rm max}$ range. Our best fitting 
values for $r_0$ and $\gamma$ are those indicated by the 
dashed-blue line (see Table 1), which corresponds to $\pi_{\rm max}=10~h^{-1}$ Mpc.

Our main clustering results, based on the joint X-ray AGN sample, are
shown in Fig.~\ref{fig:3}. The upper panel presents the projected correlation 
function, while the lower panel presents the redshift-space
correlation function (circular points) as well as the inferred
real-space correlation function, $\xi(r_p)$, (magenta
pentagons). The filled circular points are those which have been used
to fit the power-law model correlation function, which is shown as a
black continuous line (the dashed red line is the 
power-law fit for a fixed slope $\gamma=1.8$). The corresponding best
fit values for the slope $\gamma$ and the correlation length are shown
in Table 1. The uncertainty of $r_0$, indicated in Table 1, does not
include the effects of cosmic variance, which can be estimated
analytically, assuming a power-law correlation
function, according to: $\sigma^2_{cv}\simeq J_2(\gamma) (r_0/r)^\gamma$, with
$J_2(\gamma)=72/[(3-\gamma) (4-\gamma) (6-\gamma) 2^{\gamma}]$
\citep{Peebles1980}. We find $\sigma_{cv}\simeq 0.3$, which is smaller
although of the same order as the fitting uncertainty indicated in
Table 1.

We can single out three important results from the clustering analysis 
of the joint X-ray point-source sample:

\begin{enumerate} \renewcommand{\theenumi}{(\arabic{enumi})}

\item[(i)] the inferred spatial correlation length, from the power-law model fit to Eq.~ (6) or
directly from Eq.~ (7), is $\sim 20-25\%$ larger than that estimated 
from the projected correlation function, a fact which is attributed to
the correlation function underestimation imposed by 
the necessary cutoff along the $\pi$ direction in the projected 
correlation function measure, 

\item[(ii)] it appears that redshift-space distortions do not significantly
affect the $z$-space correlation function, since $\xi(s) \simeq
\xi(r_p)$, and
 
\item[(iii)] the real-space correlation function, at a median redshift of
0.976,  has a slope $\gamma\simeq 1.5$ and a correlation
length $r_0\simeq 7.2~h^{-1}$ Mpc.
\end{enumerate}
\noindent

\begin{table}
\caption{Clustering results for the joint sample of all 5 fields (1466
  sources). The clustering length units are $h^{-1}$
  Mpc. The results correspond to $\pi_{\rm max}=10~h^{-1}$ Mpc, but
  they are very similar over the whole indicated $\pi_{\rm max}$
  range, but more so for $5 \mincir \pi_{\rm max} \mincir 16
 ~h^{-1}$ Mpc, as can be seen from Fig.~\ref{fig:pmax}. The cosmic
  variance is estimated to give a further $\sim 0.3$ uncertainty in
  $r_0$.}
\label{tab_sample}
%\begin{center} 
%\scriptsize
\tabcolsep 14pt
\begin{tabular}{l c c  c } 
%\multicolumn{4}{c}{1466 sources}\\ 
\hline  
  & $\gamma$  &  $r_0$  & $r_0$ ($\gamma=1.8$)  \\  \hline  
$w_p(r_p)$ & 1.84$\pm0.07$ & 5.2$\pm0.5$ & $5.2\pm0.5$ \\
$\xi(s)$ & 1.49$\pm 0.20$ & 6.6$\pm1.0$ & $6.0\pm0.8$ \\
$\xi(r_p)$ & 1.48$\pm 0.12$ & 7.2$\pm0.6$ & $6.5\pm0.4$  \\ \hline
\end{tabular} 
%\begin{list}{}{}
%\item 
%\end{list}
%\end{center}
\end{table}

\begin{figure}
\begin{center}
\includegraphics[width=9.9cm]{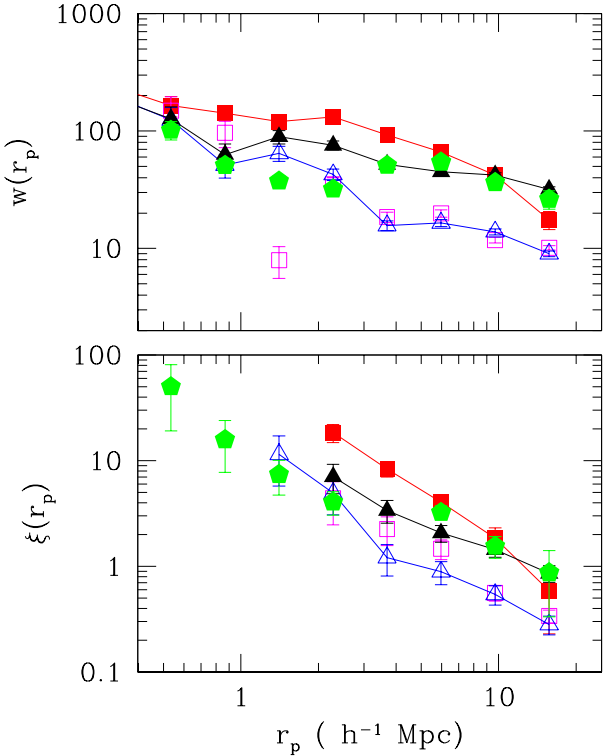} %{allart.pdf}
\end{center}
\caption{Correlation function of the individual X-ray surveys:
CDF-N (green pentagons), CDF-S (red filled squares), eCDF-S (black
filled triangles), COSMOS (magenta empty squares), AEGIS
(blue empty triangles) for redshift interval $0<z<3$). {\em Upper Panel:} Projected
correlation function. {\em Lower Panel:} The resulting spatial
correlation function.}
\label{all_f}
\end{figure}

 \subsection{The Individual Field Correlation functions}
We now perform the clustering analysis from the previous section
in each individual field and compare our results to those from the original studies.
To this end we use the redshift interval $0<z<3$ and sources with
luminosities $L_x\ge 10^{41}$ erg s$^{-1}$, in order to avoid the
gross contamination of our X-ray AGN sample by normal galaxies. 
We estimate the projected and $z$-space correlation functions in ten logarithmic
intervals of width $\delta\log_{10}(r_p,\pi)=0.175$ and
$\delta\log_{10}(s)=0.2$, respectively. With the exception of the CDF-N
field (see below), the power-law model fit was performed for
$r_p\gtrsim1.5~h^{-1}$ Mpc in order to avoid the non-linear
contributions to the correlation function. 

\begin{table*}
\caption{Parameters of the power-law fits to the different survey correlation
  functions. The results of the projected correlation function,
  $w_p(r_p)$, are shown in columns 4 and 5 and of the inferred 
  real-space correlation function, $\xi(r_p)$, in columns 6 and
  7. Finally column 8 shows the correlation length of $\xi(r_p)$ for
  the average value of the slope, $\gamma=1.44$ (which is practically
  equal to the slope of the joint sample, see Table 1).
In all cases, except for the CDF-N, the power-law fits have been applied to the
correlation function data, excluding the non-linear regime ($\mincir 1.5~h^{-1}$ Mpc). 
Since the CDF-N $w_p(r_p)$ shows a knee
around $r_p\sim3.5~h^{-1}$ Mpc, we present power-law fits for data
below and above this scale. Correlation lengths and $\pi_{\rm max}$
are in $h^{-1}$ Mpc.}\label{tab_ae0}
\tabcolsep 12pt
\begin{tabular}{l c c c c c c c c}
Survey&$\#$& $\pi_{\rm max}$&$\gamma$& $r_0$   & $\gamma_{c}$  & $r_{0,c}$& $r^{\gamma=1.44}_{0,c}$\\  \hline
CDF-N$^1$&243 & 20 &$1.72\pm0.30$&$4.6\pm0.7$ & $1.66\pm0.16$&5.1$\pm0.6$ & 6.0$\pm0.8$\\  
CDF-N$^2$&243 & 20 &$1.80\pm0.23$&$9.9\pm3.9$ & $1.46\pm0.22$& 13.3$\pm4.3$& 13.3$\pm4.3$\\  
CDF-S    &219 & 25 &$1.90\pm0.08$&$10.6\pm1.4$& $1.72\pm0.08$&12.8$\pm 1.1$&15.1$\pm 2.1$ \\  
AEGIS    &392 & 20 &$1.59\pm0.09$&$4.3\pm0.6$ & $1.22\pm0.20$& 5.5$\pm1.0$& 5.7$\pm0.8$\\  
COSMOS   &417 & 25 &$1.65\pm0.08$&$5.5\pm0.8$ & $1.36\pm0.15$& 6.8$\pm 1.0$& 7.0$\pm 0.8$\\  
eCDF-S   &288 & 10 &$1.46\pm0.07$&$7.6\pm1.0$ & $1.23\pm0.16$& 12.4$\pm2.2$& 11.1$\pm1.8$\\  \hline
\end{tabular} 
\begin{list}{}{}
\item $^1$ $r_p\mincir 3.5~h^{-1}$ Mpc
\item $^2$ $r_p> 3.5~h^{-1}$ Mpc
\end{list}
\end{table*}

In Fig.~\ref{all_f} we show the projected (upper panel) and the
inferred via Eq.~(6) spatial
(lower panel) correlation function of the individual fields, color
coded, while in Table 2 we summarize their best power-law fit
parameters $\gamma$ and $r_0$. It is evident that the individual
determinations of the correlation function are hampered by
cosmic variance. It is also interesting to note that the results of
the largest survey (COSMOS with $\sim 1$ deg$^2$ area) are consistent
with those of the Joint sample (see Table 1).

We find small differences with respect to the
results of the original studies, which are mostly due to the slightly
different sample definitions and different choices of $\pi_{\rm max}$. 
For example, for the CDF-S field we use the $2~\rm{Msec}$ 
catalog rather than the $1~\rm{Msec}$ catalog used by \citet{Gilli05}, while for the CDF-N 
\citep{Gilli05} and for the AEGIS \citep{Coil09} samples, the
corresponding authors used sources with $L_x\ge 10^{42}$ erg s$^{-1}$.
When using the latter luminosity limit and the original value of $\pi_{\rm max}$, 
we recover the original clustering results\footnote{
Due to the fact that in the luminosity dependent analysis of the next section
we are left with a small number of sources, we
are obliged to lower the X-ray luminosity limit to $L_x=10^{41}$erg s$^{-1}$.}.
The only field for which it was not possible to fit a power-law
correlation function in the linear scales
%($r_p\magcir 1 h^{-1}$ Mpc), 
is the CDF-N field, which presents a knee at
$r_p\sim 3.5~h^{-1}$ Mpc. Therefore, we have performed a power-law fit
in the two ranges where such is exhibited (i.e., for $r_p\mincir 3.5$
and $>3.5~h^{-1}$ Mpc, respectively). 

The largest clustering amplitude is observed for the CDF-S, a fact which
has been attributed to the
existence of a large supercluster at $z\simeq 0.7$ \citep{Gilli03}, 
and which appears to affect also the corresponding amplitude of
the eCDF-S field, the correlation function of which is estimated here for
the first time. However, inspecting Fig.~\ref{all_f} and Table 2, it is interesting to note
that while the small-scale ($r_p\mincir 3.5~h^{-1}$ Mpc) correlation function of CDF-N
is consistent with that of both the COSMOS and AEGIS fields, its
large-scale correlation function ($r_p> 3.5~h^{-1}$ Mpc) seems to be consistent
with that of the CDF-S and eCDF-S fields. This suggests
that either there is some unique structure affecting also the clustering 
pattern in the CDF-N (although at a lesser extent than in the CDF-S)
or there is some unknown systematic affecting both CDF's. 

 \subsection{Luminosity Dependence of Clustering}
We now wish to investigate the possibility of an X-ray luminosity
dependent AGN clustering, as suggested by \citet{Plionis08}. 
Interestingly, \cite{Cappelluti10}, have found a dependence of clustering 
on the X-ray luminosity in the sample of local ($\langle
z\rangle=0.05$) X-ray selected AGN from the SWIFT/BAT survey. Similarly,
\cite{Krumpe10} have also found such indications in the X-ray AGN sample
 from the {\it ROSAT}-All-sky-survey.

The five X-ray surveys that we have analysed in the previous sections, 
although they cover a similar redshift range, have X-ray luminosity
distributions which are quite different. To disentangle the luminosity and redshift
dependence of clustering, we construct a low-$L_x$ and a high-$L_x$
X-ray luminosity sub-sample for each survey. We also impose them to
have the same redshift distribution.
This is achieved by first splitting the sources according to the selected 
luminosity limit and then matching their binned redshift distribution in the
common area. We do so by randomly selecting sources from the most abundant
$L_x$ subsample so as to reproduce exactly the shape of the binned redshift
distribution of the less abundant subsample. The bin size used is $\delta z=
0.05$. This random selection process is performed eight times and the final
results are the average over these eight realizations.

\begin{figure}
%\begin{center}
%\includegraphics[height=0.9\columnwidth]{aegisz3.pdf}
%\includegraphics[height=0.9\columnwidth]{aegislum41.pdf}
\includegraphics[width=8.3cm]{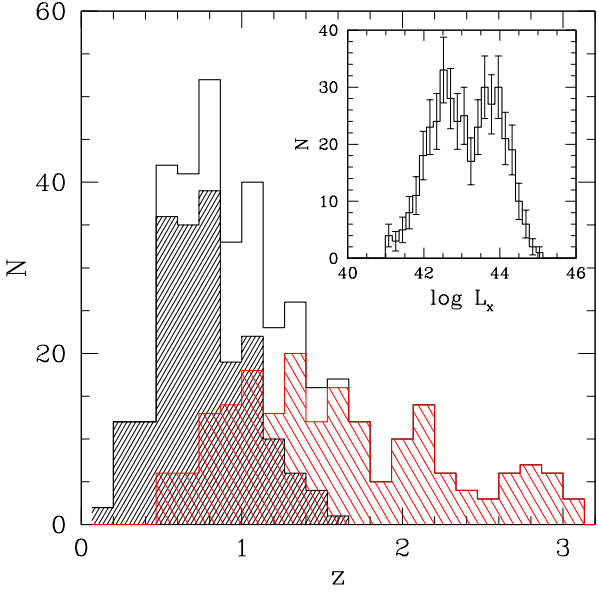}
%\end{center}
\caption{The redshift distribution of the AEGIS sources (thick black line
  histogram) divided also in the two luminosity subsamples (low-$L_x$:
  black dense-shaded histogram, and high-$L_x$: red sparse-shaded histogram).
The inset panel shows the X-ray luminosity distribution. The trough
between the two apparent peaks, at $L_x\simeq 10^{43}$ erg s$^{-1}$,
defines the division value between the low-$L_x$ and high-$L_x$
subsamples.}\label{ae_z}
\end{figure}

\begin{figure}
\begin{center}
\includegraphics[width=8.3cm]{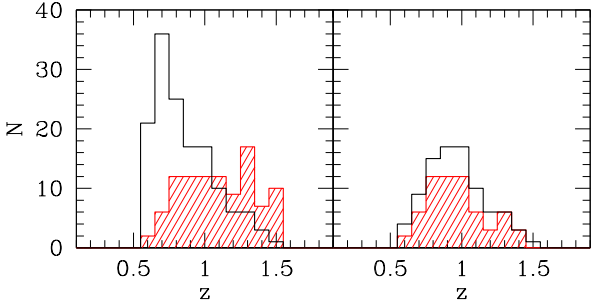}
\end{center}
\caption{{\em Left Panel:} Redshift distribution at the common 
  interval $z=0.5-1.5$ of the Low-$L_x$ (black line) and High-$L_x$ (red
  shade) subsamples. {\em Right Panel:} The matched redshift
  distributions of the further reduced subsamples, according to method described in
  the text imposed in order to cancel redshift dependent effects in
  their correlation function comparison.}
\label{faahl_z}
\end{figure}

As an example of our procedure we show in Fig.~\ref{ae_z} the redshift distribution 
%and luminosity distribution
of the whole AEGIS field (black thick line), dividing it also in its
two luminosity subsamples, i.e., the low-$L_x$ with $L_x<10^{43}$ erg s$^{-1}$
(black dense-shaded histogram) and a high-$L_x$ with $L_x\ge 10^{43}$ erg
s$^{-1}$ (red sparse-shaded histogram). In the inset panel we also show
the X-ray luminosity distribution of the AEGIS field, were a clear
division can be observed at $L_x\simeq 10^{43}$ erg s$^{-1}$.
Finally, the left panel of Fig.~\ref{faahl_z} shows the redshift
distribution of the low-$L_x$ and high-$L_x$ subsamples in the common redshift range
($0.5<z<1.5$), while in the right panel we present one random realization of the
matched $z$-distributions of the two subsamples, which are those
finally used to estimate the correlation function of the low-$L_x$
and high-$L_x$ subsamples. It is evident that our procedure has
managed to create subsamples that have a common redshift distribution,
therefore surpressing any redshift dependent effect in the comparison
of the clustering results of the low and high-$L_x$ AGN subsamples.
%{\color{red} Maybe there's room for some discussion here.
%There're two important figures shown here but no commenting on the results whatosever.} 
\begin{table}
\caption{Correlation function results from the X-ray luminosity separated
  subsamples for each individual X-ray survey. 
  Note that due to the small samples involved we have used
  different bin sizes for each sample, in order to minimize the
  intrinsic scatter. Note also that the 
  following redshift intervals were used: CDF-N $0.5<z<1.2$; CDF-S $0.7<z<1.4$;
  AEGIS $0.5<z<1.5$; COSMOS $0.6<z<1.8$; eCDF-S $0.7<z<1.6$. The last
  two columns list the slope (averaged over the two
  luminosity sub-samples) and amplitude of the power-law fit to the
  spatial correlation function, $\xi(r_p)$, of Eq.~(6).}
\label{tab_ae}
\tabcolsep 2.5pt
\begin{center} 
%\scriptsize
\begin{tabular}{l c c c c c c c c}
$\log_{10} L_x$ & $\#$ & $\bar{z}$ & $\pi_{\rm max}$ &$\gamma$  &  $r_0$ & $\gamma_c$ &$r_{0,c}$  \\  \hline
  \multicolumn{7}{c}{CDF-N}\\  
$\le 42.2$& 73& 0.84 &10 & $1.74\pm0.15$ & $3.4\pm0.4$& 1.36 &$4.7\pm0.7$ \\  
$>42.2$   & 79& 0.96 &10 & $1.42\pm0.16$ & $4.5\pm1.1$& 1.36 &$7.2\pm 1.1$\\ 
\hline  
\multicolumn{7}{c}{CDF-S}\\ 
$\le 42.5$&53& 0.95 &10 &$1.97\pm0.20$  & $10.0\pm2.5$& 1.54 &$16.5\pm2.3$ \\  
$>42.5$   &44& 1.02 &10 &$1.65\pm0.15$  & $9.3\pm2.3$ & 1.54 &$13.6\pm2.6$\\ 
\hline  
\multicolumn{7}{c}{AEGIS}\\ 
$\le 43$&90&0.83 & 20&$2.03\pm0.29$ & $4.6\pm1.5$ & 1.83 &$5.1\pm1.3$ \\  
$>43$   &63&0.82 & 20&$1.90\pm0.30$ & $5.7\pm2.1$ & 1.83 &$6.4\pm1.8$\\ 
\hline  
\multicolumn{7}{c}{COSMOS}\\ 
$\le 43.5$&72&1.06 & 30& $2.00\pm0.13$ & $6.6\pm1.3$ & 1.85 &$7.6\pm1.5$\\  
$>43.5$   &51&1.07 & 30& $2.00\pm0.23$ & $10.3\pm3.0$& 1.85 &$11.9\pm3.0$\\ 
\hline  
\multicolumn{7}{c}{eCDF-S}\\ 
$\le 43.2$&51&1.04 &30 & $1.91\pm0.36$ & $5.9\pm2.5$ & 1.85 &$6.3\pm2.5$\\  
$>43.2$   &30&1.02 &30 & $1.98\pm0.28$ & $7.7\pm3.0$ & 1.85 &$8.6\pm3.0$\\  \hline
\end{tabular} 
\begin{list}{}{}
\item 

\end{list}
\end{center}
\end{table}

We then use the $\chi^2$ minimization procedure to fit a power-law
model to the measured $w_p(r_p)$ of the low-$L_x$ and high-$L_x$
sub-samples, typically in the projected
separation range $r_p\in [0.16- 10]~h^{-1}$ Mpc. Note that it is not
possible to exclude in this analysis the non-linear scales ($\mincir
1.5~h^{-1}$ Mpc) due to the small number of sources involved 
and the consequent noisy correlation function.
Furthermore, and in
order to be able to compare the different results on an equal footing,
we present the inferred spatial correlation length, based on a
power-law fit to the $\xi(r_p)$ (Eq.~6) for a common slope, which is
the mean of the low and high-$L_x$ sub-sample $\xi(r_p)$ slopes.

\begin{figure}
\begin{center}
\includegraphics[width=8.1cm]{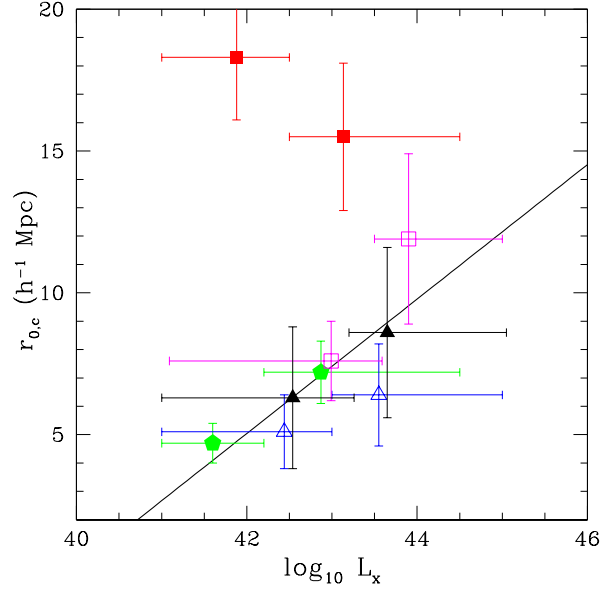}
\end{center}
\caption{Spatial clustering length, $r_{0,c}$, estimated from Eq.~(6), 
as a function of X-ray luminosity for the
different surveys: The color code and symbols are the same as in Fig.~\ref{all_f}. The
black line is the best linear fit to the data, excluding the CDF-S
(red circles) results.}\label{lm_e}
\end{figure}

The results of the correlation function analysis for the low and 
high-$L_x$ subsamples of the individual X-ray surveys can be 
found in Table \ref{tab_ae}. In Fig.~\ref{lm_e} we
present the inferred spatial clustering length, $r_{0,c}$, as a function
of the X-ray luminosity. There are indications of a luminosity
dependent clustering in all X-ray surveys, except for the CDF-S in
which the correlation lengths for both low and high-$L_x$ subsamples are
atypically very high \citep[probably due to the dominance of a known
supercluster,][]{Gilli03}. Although the luminosity dependent
clustering indications in each of the four remaining X-ray field are
rather weak, there is an overall suggestive and systematic trend with the high $L_x$
subsamples being more strongly clustered than the lower $L_x$
ones. A linear least square fit to the data of Fig.~\ref{lm_e}, taking
into account the uncertainties in both axes and excluding 
the CDF-S, provides the following dependence
$r_{0,c}/(h^{-1} {\rm Mpc})\simeq 2.4 \log_{10} L_x-94.7$. This 
is indicated by the thick continuous line in Fig.~7.
The relatively strong trend implied by this relation, although the
individual survey luminosity-dependence results are not that
significant, is due to the fact that in each field a different high
and low luminosity range is sampled, increasing the X-ray luminosity
dynamical range.

\section{BIAS AND HOST HALO MASS}
It is well established that the extragalactic sources are biased tracers
of the underlying mass fluctuation field \citep[e.g.,][]{Kaiser87,Bardeen86}.
The parameter that encapsulates this fact is the so-called linear bias
factor, $b$, defined as the ratio of the
fluctuations of some mass tracer, here the AGN
($\delta_{\rm AGN}$), to those of the underlying DM mass
($\delta_{\rm DM}$),
\begin{equation}
\label{bias}		
b=\frac{\delta_{\rm AGN}}{\delta_{\rm DM}} \;.
\end{equation}
Since the correlation function is generally defined as  
$\xi(r)=\langle \delta({\bf x}) \delta({\bf x}+{\bf r})\rangle$ (the
so-called {\em Poisson-process} definition),
an equivalent definition of the bias parameter is
the square root of the ratio of
the two-point correlation function of AGN to that of the underlying mass,
\begin{equation}\label{bias1}		
b=\left(\frac{\xi_{\rm AGN}}{\xi_{\rm DM}}\right)^{1/2}.
\end{equation}
Another related definition, which is the one that will be
used in the current work, is as the ratio of the variances of the AGN
and underlying mass density fields, smoothed at some linear scale,
%traditionally taken to be $8 \; h^{-1}$ Mpc (at which scale the
%variance is of order unity):
\begin{equation}\label{bias2}
b=\frac{\sigma_{8,\rm AGN}}{\sigma_{8, \rm DM}};,
\end{equation}
where $\sigma_{8,\rm AGN}$ is the rms fluctuations of the AGN density
distribution within spheres of a co-moving radius of $8~h^{-1}$
Mpc, given under the assumption of power-law correlations, by \cite{Peebles1980},
\begin{equation}\label{bias3}
\sigma_{8,\rm
  AGN}=J_2(\gamma)^{\frac{1}{2}}\left(\frac{r_0}{8}\right)^{\frac{\gamma}{2}}\;,
\end{equation}
% where $J_2(\gamma)=72/[(3-\gamma)(4-\gamma)(6-\gamma)2^{\gamma}$,
and $\sigma_{8,\rm DM}$ is the variance of the DM density fluctuation
field which evolves according to (e.g., Peebles 1993):
\begin{equation}\label{bias4}
\sigma_{8, \rm DM}(z)=\sigma_8\frac{D(z)}{D(0)}\;,
\end{equation}
with $D(z)$ the linear growth factor, %scaled to unity at present
                                %time, 
which for the concordance $\Lambda$ cosmology is given by (Peebles 1993)
\begin{equation}
D(z)=\frac{5\Omega_mE(z)}{2}\int_{z}^{\infty}\frac{(1+y)}{E^3(y)}\mathrm{d}y \;.
\end{equation}
Combining Eqs.~(\ref{bias2}), (\ref{bias3}), (\ref{bias4}), we obtain
the cosmological evolution of biasing as a function of the power-law
clustering parameters,
\begin{equation}
b(z)=\left(\frac{r_0}{8}\right)^{\frac{\gamma}{2}} J_2^{\frac{1}{2}}
\left(\frac{\sigma_8D(z)}{D(0)}\right)^{-1}.
\end{equation} 
Finally, an alternative approach to estimate the bias, free of the power-law
clustering assumption, is provided by the square-root of the ratio of
the projected AGN correlation function to the two-halo
term of the theoretical projected correlation function, based on
the Fourier transform of the linear CDM power-spectrum (eg., Allevato et al. 2011;
Krumpe et al. 2012). These authors have used both of the latter two
apporaches to estimate the bias and have shown that they provide
mutually consistent results.

%The best fit parameters ro and {\rm$\gamma$} are estimated fitting the
%projected correlation function {\rm$r_p=0.4-40Mpch^{-1}$}
The CDM structure formation scenario predicts that the bias factor is
determined by the mass of the DM halo within which the
extragalactic mass tracer forms \citep[e.g.,][]{Mo96}. In order
to assign a characteristic DM halo mass to the estimated bias factors 
we use two bias evolution models, the \cite{Tinker10} (hereafter
TRK) and the \cite{Basilakos2008} model (hereafter BPR). The former,
an improvement of the original \cite{Sheth99} model, belongs to the so-called 
{\em galaxy merging} bias family, which is based on the \cite{Press74} 
formalism, the peak-background split \citep{Bardeen86} and
the spherical collapse model \citep[e.g.,][]{Sheth99,Valageas09,Valageas11}.  
The latter is an extension of the so-called
{\em galaxy conserving} bias models, and uses linear perturbation theory, 
and the Friedmann-Lemaitre solutions to derive a second-order
differential equation for the evolution of bias, assuming
that the tracer and the underlying mass share the same
dynamics. Furthermore, the model includes the
contribution of an evolving DM halo population, due to processes like merging. 
Details of these models, as well as a comparison among five current
bias evolution models can be found in \citet{Papageorgiou12}.
% Using  equations (A8), (A9) and (A10) from van den Bosch et al 2002,
% assuming {\rm$ \delta_{cr}=1.69$} and {\rm$\Gamma=\Omega_mh=0.21$} we
% derived typical DM halo masses.

\subsection{X-ray AGN Bias}
\begin{table}
\caption{The bias corresponding to the clustering of the Joint sample
  (1466 sources) and of the luminosity
divided subsamples for each individual X-ray survey studied. The corresponding
values of the halo mass, in units of $\log_{10}(M_{\rm h}/[h^{-1} M_{\odot}])$,
are estimated using a WMAP7 cosmology, i.e.,
$\sigma_8=0.81$ and $\Omega_m=0.273$ and are based on the BPR (4th
column) and TRK (5th column) bias evolution model.% (see also
                                % Papageorgiou et al. 2012).
}\label{bb}
\begin{center} 
%\scriptsize
\tabcolsep 4pt
\begin{tabular}{l c c c c c }\hline           
$\log_{10} L_x$ & $\overline{\log_{10} L_x}$ &$\bar{z}$ & $b(\bar{z})$ & BPR & TRK \\ \hline
  \multicolumn{5}{c}{JOINT}\\  
           & & 0.98   &$2.26\pm0.16$ &$13.06^{+0.12}_{-0.14}$ & $13.19^{+0.07}_{-0.09}$ \\
  \multicolumn{5}{c}{CDF-N}\\  
$\le 42.2$ &41.84&0.84 &$1.52\pm 0.17$ &$12.38^{+0.29}_{-0.40}$ &$12.26^{+0.23}_{-0.30}$ \\
$>42.2$    &43.18&0.96 &$2.15\pm 0.24$ &$12.99^{+0.20}_{-0.24}$ &$12.79^{+0.19}_{-0.21}$ \\ \hline
  \multicolumn{5}{c}{CDF-S}\\  
$\le 42.5$&42.19 &0.95 &$4.22\pm 0.47$ &$14.07^{+0.15}_{-0.17}$ &$13.86^{+0.11}_{-0.12}$ \\
$> 42.5$  &43.45 &1.02 &$3.75\pm 0.56$ &$13.80^{+0.20}_{-0.25}$ &$13.52^{+0.17}_{-0.21}$ \\ \hline
  \multicolumn{5}{c}{AEGIS}\\  
$\le 43$&42.75 &0.89 &$1.74\pm 0.41$ &$12.65^{+0.45}_{-0.84}$ &$12.49^{+0.39}_{-0.64}$ \\
$> 43$ &43.86 &0.89 &$2.14\pm 0.55$ &$13.09^{+0.41}_{-0.68}$ &$12.87^{+0.35}_{-0.57}$ \\ \hline
  \multicolumn{5}{c}{COSMOS}\\  
$\le 43.5$&43.30&1.07 &$2.72\pm 0.50$ &$13.24^{+0.27}_{-0.36}$ &$13.04^{+0.24}_{-0.33}$ \\
$> 43.5$  &44.21&1.07 &$4.12\pm 0.96$ &$13.86^{+0.29}_{-0.39}$ &$13.59^{+0.24}_{-0.34}$ \\ \hline
  \multicolumn{5}{c}{eCDF-S}\\  
$\le 43.2$ &42.86&1.03 &$2.25\pm 0.83$ &$12.96^{+0.54}_{-1.16}$&$12.79^{+0.47}_{-0.97}$ \\
$> 43.2$   &43.97&1.03 &$3.00\pm 0.97$ &$13.46^{+0.42}_{-0.70}$ &$13.22^{+0.37}_{-0.62}$ \\ \hline
\end{tabular} 
\begin{list}{}{}
\item 
\end{list}
\end{center} 
\end{table}

When applying the above analysis to the clustering results of our 
Joint X-ray sample (with a median redshift $\langle z \rangle
\simeq 0.98$), we find a bias factor of $b\simeq 2.26\pm 0.16$. 
This value corresponds to a DM halo mass
of $\log_{10} (M_{\rm h}/h^{-1} M_{\odot}) \simeq 13.06 (\pm 0.13)$ and
13.19$(\pm 0.08)$ for the BPR and TRK models, respectively
(the first raw of Table \ref{bb}). 
We also extend the same analysis %of the redshift evolution of bias
to the low-$L_x$ and high-$L_x$ AGN subsamples in each X-ray field
separately. As expected by the dependence 
of the correlation function on luminosity that we presented in Section~4.3,
the high-$L_x$ subsamples provide larger bias factors and correspondingly
larger DM halo masses (for both the BPR and TRK bias models)
than those of the low-$L_x$ subsamples. 
An exception is  
the CDF-S field, which as we have already discussed, is hampered
by the presence of a large supercluster; \cite{Gilli03}. 
Thus, the data seem to suggest that high-$L_x$ sources 
inhabit more massive DM halos than low-$L_x$ sources.
The results listed in Table \ref{bb} indicate that 
for an average luminosity difference of $\langle
\delta\log_{10}L_x\rangle \simeq 1$ between the high and low-$L_x$
AGN subsamples, the corresponding average DM halo mass of the former
sample is a factor of $\sim 3$ larger than that of the latter.

Similar results have been found by \citet{Cappelluti10}, for a
sample of X-ray selected AGN from the SWIFT/BAT survey. The authors
derive for AGN with $\rm L_x<10^{44}\rm{~erg~s}^{-1}$ a DM host halo mass of
$\log_{10} (M_{\rm h}/[h^{-1} M_{\odot}])\approx 10^{12} M_\odot$, while
for AGN with $L_x\ge 10^{44}\rm{~erg~s}^{-1}$ they find
a DM host halo mass of $\sim 10^{14}~h^{-1} M_\odot$.
\cite{Krumpe10} also find evidence for luminosity dependence in
a sample of the {\it ROSAT}-All-sky-survey AGN. In particular, they find $\log_{10}
(M_{\rm h}/[ h^{-1} M_{\odot}])=11.83$ and $13.10$ for their low
and high X-ray luminosity AGN sources (below and above
$L_{0.1-2.4{\rm keV}}=10^{44.3}\rm{~erg~s}^{-1}$), respectively.

\subsection{Bias Evolution}
\begin{table}
\caption{The values of the X-ray AGN bias factor in redshift
  bins. Above $z=2$ the correlation function was extremely noisy and 
impossible to determine. The halo mass shown are estimated using a
WMAP7 cosmology, i.e., $\sigma_8=0.81$ and $\Omega_m=0.273$ and for
three different models of bias evolution
(see also Papageorgiou et al. 2012). }\label{bi_t}
\tabcolsep 4pt
\begin{tabular}{l c c c c c c} \hline 
$\#$ & $z$-range  &$\bar{z}$&
$b(\bar{z})$&\multicolumn{2}{c}{$\log_{10} (M_{\rm~h}/h^{-1} M_{\odot})$}\\ 
\multicolumn{4}{c}{} & BPR & TRK \\ \hline
%1466 &  all       &0.976   &$2.26\pm0.16$ &$13.06^{+0.12}_{-0.14}$ &$13.19^{+0.07}_{-0.09}$\\ \\
353  & 0.00 - 0.67&0.488   &$1.71\pm0.23$  &$13.37^{+0.27}_{-0.36}$ &$13.00^{+0.21}_{-0.29}$ \\
354  & 0.67 - 0.96&0.787   &$2.46\pm0.25$  &$13.52^{+0.16}_{-0.19}$ &$13.22^{+0.14}_{-0.16}$\\
352  & 0.96 - 1.38&1.123   &$2.46\pm0.23$  &$13.00^{+0.16}_{-0.18}$ &$12.90^{+0.10}_{-0.22}$ \\
200  & 1.38 - 2.00&1.613   &$3.85\pm1.55$  &$13.09^{+0.48}_{-0.88}$ &$13.00^{+0.43}_{-0.82}$ 
\\ \hline
\end{tabular} 
\begin{list}{}{}
\item 
\end{list}
\end{table}

\begin{figure*}
\includegraphics[width=8.3cm]{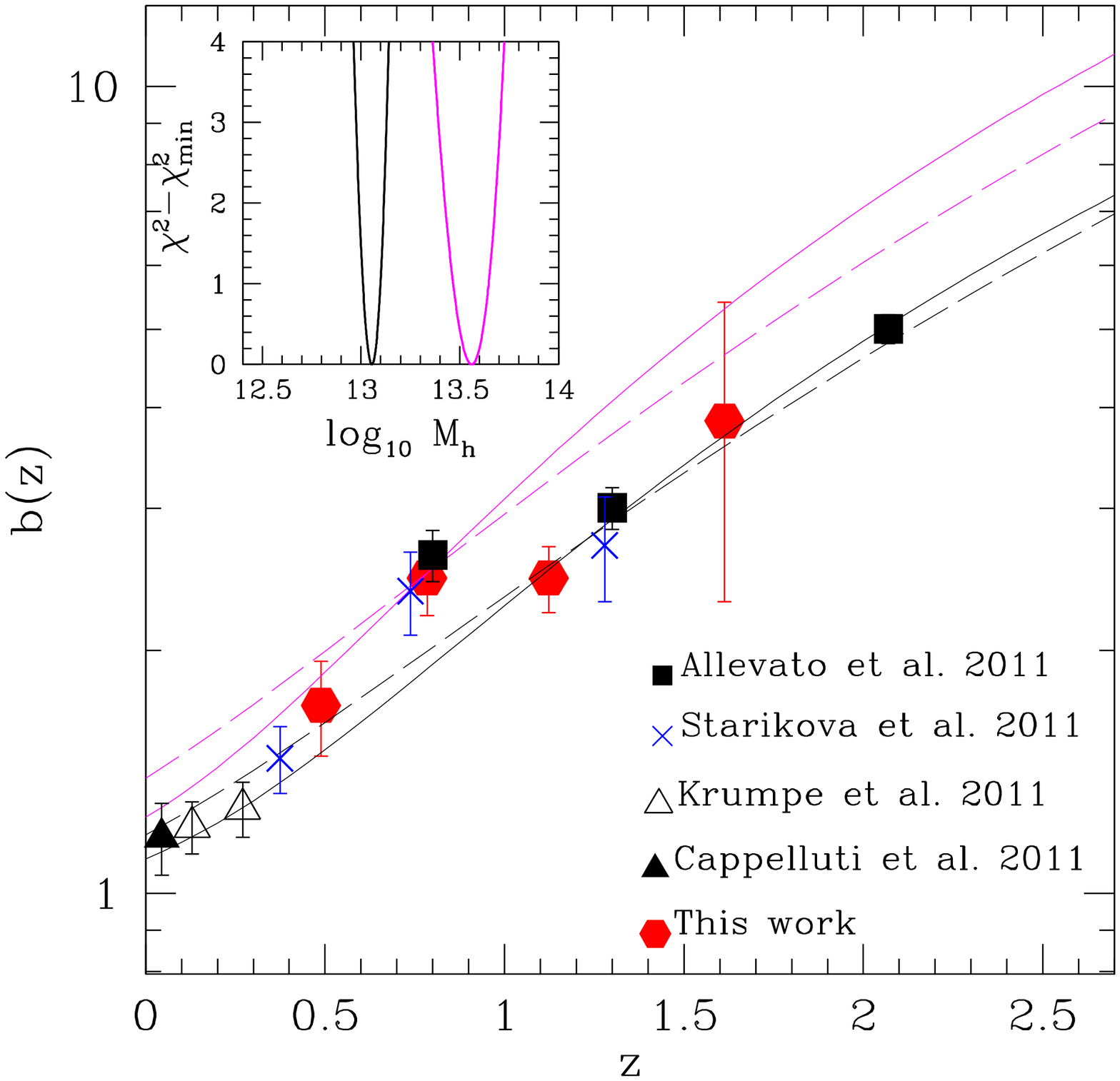} \hfill
\includegraphics[width=8.3cm]{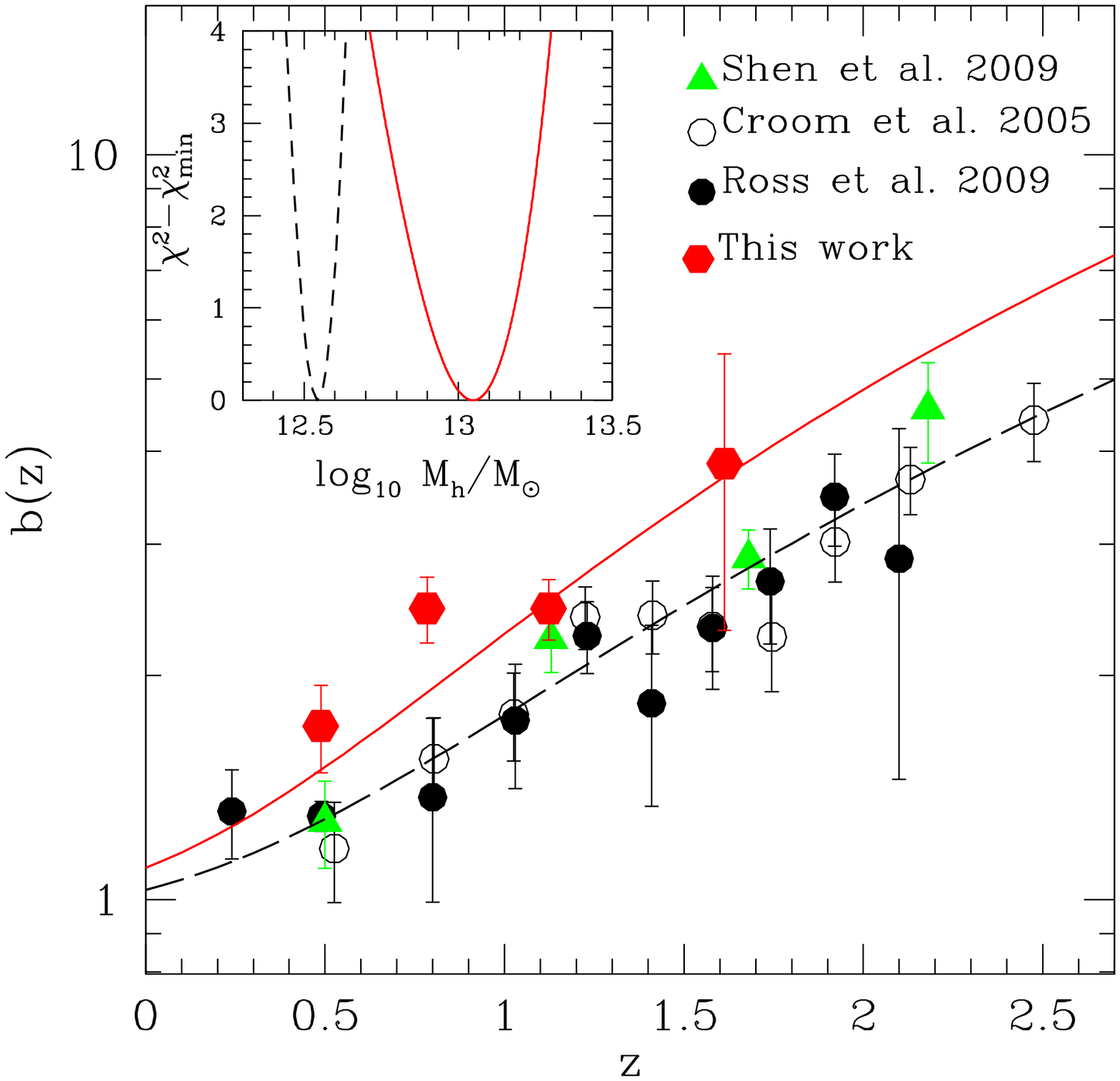}
\caption{
{\em Left Panel:} 
Comparison of the redshift evolution of the bias parameter of our Joint X-ray
sample (red pentagons), with that of other X-ray surveys. The point
and color coding of the results based on different surveys is
indicated by the corresponding labels.
%\citet{Allevato11} - filled squares; \citet{Starikova11} - blue crosses; 
%\citet{Cappelluti10} - filled triangle, and \citet{Krumpe10} open
%triangles. 
The theoretical bias evolution models, fitted to the data for a single DM halo mass,
are shown as continuous or dashed lines: 
The black continuous line is the best fit BPR
bias evolution model to all the bias data, excluding the deviant
points at $z\sim 0.75$, while the dashed
line is the corresponding best fit TRK model. 
The magenta continuous and dashed lines are the fits of the
corresponding two models to the bias values around $z\sim 0.75$. 
In the inset panel we show the best fit DM halo mass value as a function of
$\chi^2-\chi_{min}^2$ for the BPR model. A value of $\chi^2-\chi_{min}^2=1$ or 4
correspond to the 1 and 2$\sigma$ uncertainty in the fitted parameter
(for the case of one free parameter).
{\em Right Panel:} Comparison of the redshift evolution of the bias parameter of the Joint X-ray
  sample (red pentagons), to that of optical QSOs. 
The point and color coding of the results based on different surveys is
indicated by the corresponding labels.
%\citet{Croom05} - open circles; \citet{Ross09} - filled circles;
%\citet{Krumpe10} - filled squares; Ross et al. 2009 - filled
%triangles; Shen et al. 2010 - filled triangles) 
The curves correspond to the BPR bias evolution model 
fitted to the X-ray data (red continuous line) and to the optical QSO
(black dashed line).
Note that the plotted bias values are rescaled to the WMAP7 cosmology,
as in Papageorgiou et al. (2012).}
\label{bias_all}
\end{figure*}

In order to investigate the redshift evolution of the bias factor we
split the whole sample in four redshift bins. The resulting values 
of the bias factor in each bin, as well the typical DM halo masses, 
based on the two previously mentioned models, are shown in 
Table~\ref{bi_t}. Fig.~\ref{bias_all} (left panel) shows our
determination of the bias in the different redshift bins together with 
the corresponding results of the {\it XMM}-COSMOS 
\citep{Allevato11}, the XBOOTES \citep{Starikova11}, the 
{\it ROSAT} \citep{Krumpe10} and SWIFT/BAT (\cite{Cappelluti10}) surveys. 
The increase of the bias factor with increasing redshift
is evident, as well as the consistency of the derived bias values of
the different studies at the corresponding redshifts.
 
\cite{Allevato11} has argued that models of bias evolution based
on a single host halo mass cannot explain their results, but rather there
are indications for different halo masses hosting X-ray AGN at the
different redshifts. We check their suggestion by
applying the previously mentioned bias-evolution models,
and attempting to fit a single halo mass jointly to all the previously
mentioned bias data. Indeed no such model can fit adequately well the results (the
resulting reduced $\chi^2$ is $\magcir 1.6$). The cause of this
failure is mostly the deviant bias values around $z\sim 0.75$. If
this redshift bin is excluded, then 
the bias data can indeed be fitted by a single host halo mass with 
$\log_{10} (M_{\rm h}/[h^{-1} M_{\odot}])\simeq 13.06 \pm 0.04$, $\chi^2$/df$=0.24$
and a present epoch bias of $b(0)=1.1$ (black continuous line in
Fig.~\ref{bias_all}) for the BPR model. For the TRK model we find
$\log_{10} (M_{\rm h}/[h^{-1} M_{\odot}])\simeq 12.94 \pm 0.05$, 
$\chi^2$/df$=0.48$ and $b(0)=1.18$ (black dashed line in Fig.~\ref{bias_all}).
If we use however, only the three deviant bias values around $z\simeq 0.75$, we
find $\log_{10} (M_{\rm h}/[h^{-1} M_{\odot}])\simeq 13.56 \pm 0.09$, $\chi^2$/df$=0.29$
and $b(0)=1.24$ (magenta continuous line) for the BPR model, and 
$\log_{10} (M_{\rm h}/[h^{-1} M_{\odot}])\simeq 13.28 \pm 0.10$, $\chi^2$/df$=0.29$
and $b(0)=1.39$ (magenta dashed line) for the TRK model.
It is interesting to point out that the bias values of the current
study (red pentagons), when excluding the deviant point, provide exactly
the same %bias evolution model 
fit as the one obtained previously, using all the determinations of the 
bias evolution values (again excluding the deviant points).
However, the uncertainty is, as expected, larger 
(for example, for the BPR model we obtain $\log_{10} (M_{\rm h}/[h^{-1} M_{\odot}])
\simeq 13.08 \pm 0.14$ with a $\chi^2$/df$=0.48$).

\subsection{X-ray versus Optically selected AGN}
We also compare in the right panel of Fig.~\ref{bias_all} our
X-ray AGN results with the bias of optically selected QSO, 
based on the 2dF spectroscopic sample of 20000 QSOs \citep{Croom05},
on an SDSS (DR5) QSO sample of $\sim$30000 spectroscopic QSO with
$z\mincir 2.2$ \citep{Ross09} and
on a homogeneous sample of $\sim$38000 SDSS QSOs with $0.1\leq z\leq5$
\citep{Shen09}. Since the X-ray data do not extend to such high
redshifts and in order to consistently compare optical and X-ray bias data, we
choose to exclude the very high-$z$ (3$<z<4.5$) optical bias data of Shen et al. 
Note also that since each of the optical QSO studies have used a slightly different
flat $\Lambda$CDM background cosmology, the bias values shown in
Fig.~\ref{bias_all} have been scaled to
the WMAP7 cosmology, following the prescription of \cite{Papageorgiou12}.
The results of each of the different studies are indicated in the Figure with
distinct colours and symbol types.

The BPR model that fits simultaneously all the optical QSO bias data,
provides a DM halo mass of 
$\log_{10} (M_{\rm h}/[h^{-1} M_{\odot}]) \simeq 12.50 \pm 0.05$ with 
%$M_{\rm h}\simeq 3.2 (\pm 0.4) \times 10^{12}~h^{-1} M_{\odot}$ with
$\chi^2/$df$=0.68$. The corresponding values for the TRK bias model
are practically the same: $\log_{10} (M_{\rm h}/[h^{-1} M_{\odot}]) \simeq 12.49 \pm 0.04$ with 
%$M_{\rm h}\simeq 3.1 (\pm 0.3) \times 10^{12}~h^{-1} M_{\odot}$ with
$\chi^2/$df$=0.64$. 
It is therefore evident that the X-ray selected AGN bias values, and the corresponding
DM halo masses, are significantly larger than those of the optical QSO,
indicating that they probably constitute separate families of
AGN, each probably having a distinct BH fueling mechanism.
\cite{Allevato11} also conclude that the DM halos
in which X-ray selected AGN reside are more massive than those of optically
selected QSOs. Further below we will provide a tentative explanation
for such a difference between the host halos of optical and X-ray selected AGN.

\section{Comparison with AGN accretion models} 
Our analysis provides new insights into the host DM halos of moderate luminosity
X-ray selected AGN. By fitting two different bias evolution
 models, we find that the  AGN in our samples inhabit
 DM halos with masses of $\log_{10}
 (M_{\rm h}/[h^{-1} M_{\odot}])\simeq 13.1 \pm 0.1$. The estimated halo mass
 is  significantly higher than the typical halo mass of optically selected QSO
 [$\log_{10} (M_{\rm h}/[h^{-1} M_{\odot}])\sim 12.5$; see
 Fig.~\ref{bias_all}]. Similar results are also found in other X-ray 
studies \citep[e.g.][]{Coil09, Starikova11}, \citep{Allevato11, Mountrichas12}
and therefore the DM halos masses estimated in most X-ray surveys 
are consistent mostly with those of elliptical/red galaxies.
This implies that the fueling mechanism of the moderate
luminosity X-ray selected AGN and the more luminous optically selected
QSOs is  different. 

Semi-analytic models that assume major mergers as the main mechanism for 
triggering AGN activity predict that the mass of DM halos that host AGN
is lower than that estimated from observations of moderate luminosity
X-ray AGN \citep[e.g.,][]{Marulli09, Bonoli09}.
The theoretical predictions from these models are more consistent with 
the halo mass of optically selected QSOs.  
Additional evidence against the major mergers scenario in moderate
luminosity X-ray AGN comes from the morphological analysis of AGN in
the AEGIS and COSMOS surveys \citep{Georgakakis09, Silverman09}. 
These authors find that X-ray selected AGN span a large
range of environments and morphologies with roughly
equal numbers of bulges, spirals and morphologically disturbed
galaxies \citep[see also][]{Schawinski11, Kocevski12, Cisternas11}.
Also stochastic accretion models \citep{HopkinsHernquist06} 
cannot explain large DM halos for moderate luminosity AGN. 
According to these models, disc instabilities and minor interactions 
feed at high accretion rates relatively small black holes. 
These models predict that moderate luminosity AGN reside in 
low density environments, similar to those of blue star-forming galaxies. 

\begin{figure*}
\includegraphics[height=0.83\columnwidth]{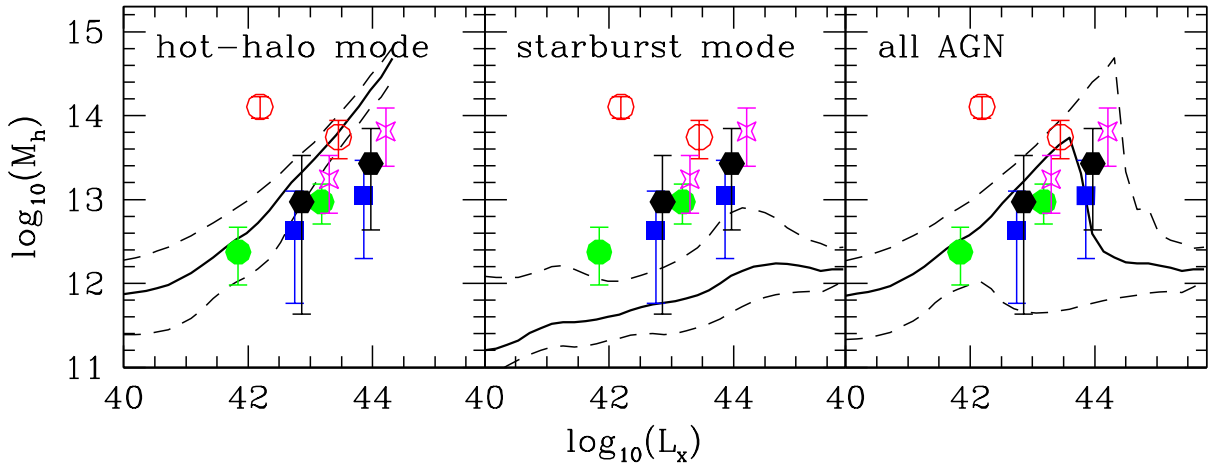}
\caption{The median X-ray luminosity (0.5-8 keV)  
of the luminosity separated subsamples (from Table~4) \emph{vs.} the DM halo mass 
predicted by the TRK bias model. The color coding is same as in Fig.~
\ref{all_f}. For comparison we show the theoretical predictions of the 
\citet{Fanidakis12} black hole model at $z=0.5-1.5$. Solid lines indicate the median of
the $\log L_{\rm x}-\log M_{\rm h}$ correlation for AGN accreting from
the hot halo  (hot-halo mode; left panel),
or during disc instabilities and mergers (starburst mode; middle panel)
and all AGN in their model (right panel). The dashed lines indicate the $10-90$ 
percentiles around the median.}
 \label{fan}
\end{figure*}

Regarding the luminosity dependence of the clustering of X-ray
selected AGN (and consequent luminosity dependence of the
 host halo mass) various theoretical models of black hole and galaxy co-evolution predict
only a weak dependence of AGN clustering on luminosity, 
\citep[e.g.,][]{Lidz06, Hopkins05, HopkinsHernquist06}.
In these models, both the bright and faint AGN reside in
similar mass halos. The bright AGN correspond to black holes that radiate at their peak 
luminosities, i.e. at accretion rates close to the Eddington ones. In
contrast, the faint end of the AGN luminosity function consists of AGN
in dimmer (late) phases in their evolution.

Our derived DM halo masses of  X-ray selected AGN 
could suggest, as \citet{Mountrichas12} also point out, a 
fueling mechanism similar to that of \citet{Ciotti01}. The 
authors suggest that  stellar winds in early-type
galaxies provide the gas supply to the black hole (see also Kauffmann
\& Heckman 2009).
Alternatively, another plausible model is that of \citet{Bower06},
where gas is accreted onto BHs
directly from the hot halo of the galaxy. \cite{Fanidakis12}, recently presented a 
calculation for accreting black holes within the \citet{Bower06} 
semi-analytic model, where AGN activity is coupled to 
the evolution of the host galaxy. The authors assume that black holes 
grow via accretion triggered by galaxy mergers 
and disk instabilities (starburst mode), as well as
accretion of hot gas from the halo of the galaxy (hot-halo mode).  
In Fig.\ref{fan},  we compare our  halo mass estimates for the individual 
subsamples in Table~5 with the predictions of the 
\cite{Fanidakis12} model for the $L_{x}-M_{h}$ correlation at $z=0.5-1.5$. 
Their calculations assume a flat cosmology with
$\Omega_m=0.25$, $\Omega_b=0.045$, $\sigma_8=0.9$ and $h=0.7$, and for
consistency, we re-estimated the DM halo mass values 
according to their adopted cosmology. 

As illustrated by the left panel of Fig.~~\ref{fan}, 
the \citeauthor{Fanidakis12} model predicts a strong X-ray luminosity -- halo mass 
correlation for AGN accreting in the hot-halo mode. In contrast, the halo mass of AGN 
in the starburst mode shows a very modest increase to X-ray luminosities 
of $L_{\rm X}\simeq10^{44}\rm{ergs^{-1}}$, beyond which it remains
constant and approximately equal to $10^{12}\rm{M_{\odot}}$ (middle panel). When considering 
all the AGN in the model, the shape of the correlation is
dominated by the hot-halo mode in the low-luminosity  and 
by the starburst mode in the high-luminosity regime
(right panel). Thus, the resulting correlation increases steeply in the moderate 
luminosity regime and then sharply declines and flattens for the brightest luminosities.
Our X-ray selected AGN results appear to be in very good agreement
with the predictions of the Fanidakis et al. model. The data lie in the 
region of the $L_{\rm X}-M_{\rm h}$ plane,
which is dominated by the hot-halo mode. This suggests that intermediate luminosity AGN 
are preferentially powered by accretion of gas from the hot halo.  
Their model further proposes that sources with $L_{\rm
  X}\gtrsim10^{44}\rm{ergs^{-1}}$, expected to be
visible as QSOs, live in haloes with masses of $\sim10^{12}M_{\odot}$ as 
suggested by clustering studies of optically selected QSOs (see Fig.~8,
right panel).

The distinctive environmental dependence of the starburst and hot-halo mode 
in the \citeauthor{Fanidakis12} model is linked to the cooling properties of the 
gas in DM haloes \citep{WF91}. In the starburst
mode, disc instabilities and galaxy mergers take place in gas rich
environments. In the standard $\Lambda$CDM paradigm, this corresponds 
to intermediate mass DM haloes, where gas cools rapidly. 
In more massive haloes, the gas is in
quasi-hydrostatic equilibrium and therefore characterized
by a significant lower cooling efficiency. In addition, in this mass regime AGN
feedback reheats the gas in the halo, shutting off the cooling completely. 
AGN activity maximizes at the transition between the rapid-cooling 
and quasi-hydrostatic regime, which in the Fanidakis et al. model is found to take place at 
$M_{\rm h}\sim10^{12}M_{\odot}$. In haloes more massive than $\sim10^{12}M_{\odot}$,
 black holes accrete low-density gas directly from the hot halo. In
 this case, the authors assume that the amount of gas that is
 accreted onto the BH is determined by, and in fact is proportional
 to the cooling luminosity of the gas in the halo
 \cite[see][]{Bower06,Fanidakis11}. Since the cooling luminosity
 increases with increasing halo mass, their model predicts a strong
 correlation between accretion luminosity and DM halo mass. 
The good agreement of the predictions of these accretion prescriptions 
with our analysis, but also with the clustering results of optically selected QSOs, 
strengthens the idea of two distinct fueling modes; the hot-halo mode in moderate
X-ray AGN and the starburst mode in the more luminous optical QSOs.

\section{SUMMARY}
We use 1466 X-ray selected AGN in the 0.5-8 keV band with
spectroscopic redshifts spanning the redshift
interval $0<z<3$, with a median of $\bar{z}=0.976$. 
We derive a spatial clustering length of $r_0= 7.2\pm 0.6
h^{-1}$ Mpc and a slope of $\gamma=1.47\pm 0.12$. The corresponding
clustering length for the nominal slope of $\gamma=1.8$ is $r_0=6.5 \pm
0.4~h^{-1}$ Mpc. The above clustering length corresponds to a bias of
$b(\bar{z})=2.26 \pm 0.16$, translating to a mass of the host DM halo of
$M_{\rm h}\sim 1.3 \times 10^{13}~h^{-1} M_\odot$.
The derived bias and the corresponding host halo mass of X-ray
selected AGN is significantly higher than that of optically selected
AGN. This may imply a different fueling mechanism
for X-ray selected AGN in comparison to optically selected AGN.
We also find indications for a dependence of the clustering strength
on the X-ray luminosity, 
when we consider AGN at a redshift of $z\sim 1$. The more X-ray
luminous the sources are the larger, typically, the clustering length and
hence the higher the DM host halo mass is.
For an average luminosity difference of $\langle
\delta\log_{10}L_x\rangle \simeq 1$, between the high and low-$L_x$
AGN subsamples, the corresponding average DM halo mass of the former
sample is a factor of $\sim 3$ larger than that of the latter.

The luminosity-dependent clustering that we find seems to be in favor of a 
black hole accretion model where moderate luminosity X-ray AGN, at
$z\sim 1$, are fueled mostly by accretion from the hot halo around  the
host galaxy. The good agreement of the model predictions with similar
studies in the optical further suggests that luminous quasar activity
is triggered in lower-mass halo (and gas-rich) environments by disk
instabilities and galaxy mergers.

\section*{Acknowledgments}
We thank Alison Coil for providing us with the spectroscopic data of
the AEGIS field, as well as for useful comments on the paper.
Part of this work was supported by the COST Action MP0905 "Black Holes
in a Violent Universe" 
\bibliography{ref}{}
\bibliographystyle{mn2e}

\end{document}